\documentclass[11pt,a4paper]{article}
\usepackage{jheppub} 
\usepackage{bm}
\usepackage{slashed}

\title{Parity violating gravitational response and  
anomalous constitutive relations}

\author[a]{Juan L. Ma\~nes}
\author[b]{and Manuel Valle}

\affiliation[a]{Departamento de F\'\i sica de la Materia Condensada, 
Universidad del Pa\'is Vasco UPV/EHU, \\
Apartado 644,  48080 Bilbao, Spain}
\affiliation[b]{Departamento de F\'\i sica Te\'orica, 
Universidad del Pa\'is Vasco UPV/EHU, \\
Apartado 644,  48080 Bilbao, Spain}

\emailAdd{wmpmapaj@lg.ehu.es}
\emailAdd{manuel.valle@ehu.es}

\abstract{
We compute the parity violating part of the time-dependent gravitational response function of an ideal gas of Weyl fermions up to third order in the derivative expansion and give its full tensorial structure. Our main results are two functions that parametrize the energy-momentum tensor in terms of gauge-invariant combinations of vector and tensor metric perturbations. The zero frequency limit of these functions is related with the anomalous constitutive relations and with the full anomalous partition function in the presence of gauge and mixed anomalies. In particular, our results imply the existence of a previously unknown invariant contribution to the parity-odd partition function at third derivative order that we explicitly construct. Beyond the static limit, the gravitational response function may provide valuable insights into time-dependent phenomena driven by anomalies.
}

\begin{document}
\maketitle
\flushbottom


\section{Introduction}

One of the most interesting and surprising developments in thermal field theory 
has been the recent realization that anomalies induce 
unexpected modifications  in the constitutive relations of relativistic 
hydrodynamics~\cite{Son:2009tf,Neiman:2010zi,Newman:2005hd}.
As a consequence, there are allowed transport  
phenomena~\cite{Kharzeev:2009pj,Kharzeev:2010gr,Landsteiner:2011cp}  related only 
to  anomalies  which manifest themselves  by parity violation.
An important feature of the new anomalous terms  is that they do not break time-reversal symmetry and hence
the equilibrium response to a perturbation is not accompanied by an increase in entropy. 
In this regard, it has been argued that a systematic derivation of constraints on 
the constitutive relations should be possible solely from the knowledge of a consistent functional 
describing the equilibrium thermodynamics,  
without having to make use of an entropy current~\cite{Jensen:2012jh,Jensen:2012jy,Banerjee:2012iz}.  
More specifically, this functional has been constructed  for an anomalous charged fluid,  
to second order in the derivative expansion in 1+1 dimensions~\cite{Valle:2012em}, 
and  to third order in 3+1 dimensions~\cite{Jensen:2012kj}. 
These constructions, supplemented by  the appropriate 
Bardeen-Zumino terms~\cite{Bardeen:1984pm,Bilal:2008qx,Jensen:2012kj} to be added to the consistent currents, 
provide higher-order anomalous contributions to the constitutive relations due to 
gravitational and mixed anomalies. 

As mentioned above, an important feature of the anomalous contributions to the constitutive relations is that they  break invariance under parity. To see the implications of this fact, 
let us  consider the non-relativistic limit of the anomalous constitutive relation connecting the vorticity 
to the momentum density~\cite{Loganayagam:2011mu,Loganayagam:2012pz}
\begin{equation}
\bm{g} = \chi_{PV} \bm{\nabla} \times \bm{v} , 
\end{equation} 
where $\bm{v}$ is the fluid velocity in a generic frame (not in the Landau frame).  
According to the  theory of hydrodynamic fluctuations~\cite{Kadanoff}, 
the non-zero static susceptibility $\chi_{PV} $ is given by the following sum rule  
\begin{equation}
\chi_{PV}  \propto \lim_{q \to 0}  \frac{i q^j  \epsilon^{j m n}}{q^2} \int_{-\infty}^\infty  d \omega\, 
 \frac{\chi_{g g}^{m n}(\omega, \bm{q})}{\omega} , 
\end{equation}
in terms of the imaginary part $\chi_{g g}^{i j}(\omega, \bm{q})$ of the  retarded momentum-momentum correlator.   
For a  translationally invariant system  this correlator takes the general form
\begin{equation}
\label{correlador}
\chi_{g g}^{i j}(\omega, \bm{q}) = \frac{q^i q^j}{q^2} \chi_L(\omega, q) + 
                              \biggl( \delta^{i j} - \frac{q^i q^j}{q^2} \biggr) \chi_T(\omega, q) + 
                               i \epsilon^{i j k} q^k \chi_{PV}(\omega, q) . 
\end{equation}

Since all the momentum density components  $g^i$  have the same signature under discrete symmetries,  
 the last term of  \eqref{correlador}  can be  different from zero only if the equilibrium state used in the definition of
 $\chi_{g g}^{i j}(t, \bm{x} ,0,\bm{0}) = \langle\left[ g^i(t, \bm{x}), g^j(0, \bm{0})\right] \rangle$ 
 is not invariant under parity.
For a system of chiral fermions this requires a nonzero  axial chemical potential,   
and $\chi_{PV}$ is  expected to be an odd function of  this potential. 
This observation suggests that the study of the parity violating piece of the two-point function for gravitons  
may be useful in order to deepen our understanding of anomalies in hydrodynamics and also to gain 
information about  time-dependent phenomena driven by anomalies, 
a field that remains relatively unexplored. 

In this paper we  undertake the computation of  the parity violating  part of  the 
gravitational response function of an ideal gas of  Weyl fermions at  small frequencies and momenta  
compared to the chemical potential and temperature. 
In other words, we  compute  the gravitational response function in the hard-dense-loop approximation. 
The procedure is based on previous work by Rebhan~\cite{Rebhan:1990yr},  who has given the 
full tensorial structure of the leading corrections $\propto T^4$ of the graviton self-energy.  
A brief review of the work by Rebhan, generalized to non-vanishing chemical potential, is presented in the next section, 
where the resulting energy-momentum tensor is shown to correspond 
to the constitutive relations of a perfect conformal fluid at lowest order in the derivative expansion of the fluid fields. 
This is extended to third order in the derivative expansion in section~3, 
where  our  main results are two functions that parametrize the energy-momentum tensor in terms 
of gauge-invariant combinations of vector and tensor metric perturbations.    
The zero frequency limit of these functions is related with the anomalous constitutive relations in section~4, 
and  with the form of the full anomalous partition function in the presence of gauge and mixed anomalies in section~5, 
where we find a previously unknown invariant contribution to the parity-odd partition function. 
Our conclusions and possible applications of this work to the study of time-dependent hydrodynamic phenomena driven by anomalies are presented in section~6.


 \section{Leading contribution at low momentum}

In this section we briefly review previous work by Rebhan~\cite{Rebhan:1990yr} on the graviton self-energy at leading order, and connect it with the constitutive relations of a perfect conformal fluid. The graviton polarization tensor is defined by
\begin{equation}\label{def}
\Pi^{\mu \nu\, \rho \sigma}(x-y) = - 4\left. \frac{\delta \Gamma}{\delta g_{\mu \nu}(x) \delta g_{\rho \sigma}(y)} \right|_{g=\eta} 
 =  -2\left. \frac{ \delta} {\delta g_{\mu \nu}(x)} \left(\sqrt{-g} \langle T^{\rho \sigma}(y)  \rangle\right)\right|_{g=\eta} , 
\end{equation} 
where $\Gamma$ is the effective action and
\begin{equation}
 \langle T^{\mu \nu}  \rangle= \frac{2}{\sqrt{-g}}\frac{\delta\Gamma} {\delta g_{\mu \nu}} .
\end{equation}
Note that eq.~\eqref{def} implies
\begin{equation}
\label{linres}
\delta\left(\sqrt{-g} \langle T^{\mu \nu}(x)\rangle\right) =  -\frac{1}{2}\int d^4 y\,  \Pi^{\mu \nu \, \rho \sigma}(x-y) 
 h_{\rho \sigma}(y) ,  
 \end{equation}
 where  the retarded version of  $\Pi^{\mu \nu \, \rho \sigma}(x-y)$ has to be  used 
 in order to compute the corresponding induced change to linear order in $h_{\mu \nu} \equiv g_{\mu \nu} - \eta_{\mu \nu}$.
  
It was  shown in~\cite{Rebhan:1990yr} that,  for small momenta $|q^0|, q \ll T_0, \mu_0$ 
the leading  behavior of $\Pi^{\mu \nu \, \rho \sigma}(Q)$ can be written in terms of  the integral  
\begin{equation}
\label{four}
I^{\mu \nu \rho \sigma}(Q) = T_0 \sum_{\omega_m} \int \frac{d^3 k}{(2\pi)^3} 
\frac{K^\mu K^\nu K^\rho K^\sigma}{K^2 (K+Q)^2}, \qquad K^\mu = (i \omega_m + \mu_0,  \bm{k}),\; \;Q^\mu = (i \nu_n, \bm{q}), 
\end{equation}
where $\nu_n$ ($\omega_n$) are bosonic (fermionic) Matsubara frequencies, 
and the pure vacuum ($T_0\!=\!\mu_0\!=\!0$) divergence has been subtracted.  
Concretely, the following combination of indices, 
\begin{equation}\label{lead}
\Pi^{\mu \nu \, \rho \sigma}(Q) = 2I^{\mu \nu \rho \sigma}(Q) -  \eta_{\alpha \beta}\left( 
I^{\alpha \beta \mu \rho} \eta^{\nu \sigma} +
I^{\alpha \beta \nu \rho} \eta^{\mu \sigma}+
I^{\alpha \beta \mu \sigma} \eta^{\nu \rho}+
I^{\alpha \beta \nu \sigma} \eta^{\mu \rho}
\right) , 
\end{equation}
produces the correct graviton polarization tensor satisfying the Ward identities implied by general covariance. 
After the sum over Matsubara frequencies is performed, the small momentum  behavior 
of $I^{\mu \nu \lambda \sigma}(Q)$ 
can be extracted  by making  the rescaling 
$q^0 = i \nu_n \to \epsilon q^0, \bm{q} \to \epsilon \bm{q}$,  and  keeping only $\mathcal{O}(\epsilon^0)$ terms. 
The integrand obtained in this way is proportional to 
the energy density which, for a left (or right)-handed Weyl field, is given by
\begin{equation}
\varepsilon=\int_0^\infty \frac{dk \, k^2}{2\pi^2} k \bigl(n_F(k-\mu_0) + n_F(k+\mu_0)\bigr) = 
\frac{7 \pi^2 T_0^4}{120}  + \frac{\mu_0^2 T_0^2}{4} +  \frac{\mu_0^4}{8 \pi^2} .
\end{equation}
The remaining angular integration yields the entire dependence  of $I^{\mu \nu \lambda\sigma }$ on $q^0/|\bm{q}|$.
In particular, the explicit computation of $I^{\mu \nu \rho \sigma}(Q)$ in the static limit ($q^0 \!=\!0$)  
yields the following  non-vanishing components~\cite{Rebhan:1990yr} for the polarization tensor
\begin{align}\label{static}
\Pi^{00 \, 00} &= -5 \varepsilon  , & 
\Pi^{00 \, i i} &= -\varepsilon , & 
\Pi^{0i \, 0i} &= -\frac{\varepsilon}{3} , \nonumber\\ 
\Pi^{i i \, j j} &= -\frac{\varepsilon}{3} , & 
\Pi^{i i \, 00} &= -\varepsilon , & 
\Pi^{i j \, i j} &= \frac{\varepsilon}{3}  ,    
\end{align}
where the value of $\Pi^{i i \, jj}$ is valid only for $i\neq j$, with $\Pi^{i i \, ii}=\varepsilon/12$.

An important feature of these results is that its form exactly corresponds to the constitutive relation  of a perfect conformal fluid 
at lowest order in the derivative expansion of the fluid fields
\begin{equation}\label{perfect}
\langle T^{\mu \nu}\rangle =  (\varepsilon + P) u^\mu u^\nu + P g^{\mu \nu} .
\end{equation}   
This can be seen by considering the system in a curved background.  
The most general static metric which is preserved by the Killing vector $\partial_t$ may 
be written as \cite{Banerjee:2012iz}
\begin{equation}\label{backg}
ds^2 = -e^{2 \sigma(\vec x)} \left(dt + a_j(\vec x) dx^j \right)^2 + g_{i j}(\vec x) dx^i dx^j . 
\end{equation}
Now, in a  comoving coordinate system,  the fluid velocity becomes  
$u^\mu = \delta_0^\mu e^{-\sigma}$, 
and  the temperature and chemical potential also acquire a dependence on the position given by~\cite{Tolman30} 
\begin{equation}
\begin{split}
T(\vec x) &= (-g_{00})^{-1/2}T_0  = e^{-\sigma} T_0, \\ 
\mu(\vec x)&= e^{-\sigma} \mu_0.  
\end{split}
\end{equation}
Here $T_0$ and $\mu_0$  are constants  which may be viewed as the 
temperature and chemical  potential in the absence of 
gravity\footnote{Note that  $T_0$ and $\mu_0$ are the same constants that appear in the  partition function 
$Z = \text{Tr}\, e^{-(H - \mu_0 N)/T_0}$ even if the Hamiltonian includes the static gravitational field.}.  
Thus, in the conformal case $P = \varepsilon/3$, the induced corrections  
to $\langle T^{\mu \nu} \rangle$  are simply obtained by making the replacements
 $\varepsilon \to  e^{-4 \sigma} \varepsilon,    
u^\nu \to e^{-\sigma}  \delta_0^\nu$  and $\eta^{\mu \nu} \to g^{\mu \nu}$. 
To linear order in $h_{\mu \nu}$  this yields 
\begin{equation}
\label{perpol}
\begin{split}
\delta \left( \sqrt{-g}  \langle T^{00} \rangle \right) &= \frac{5 \varepsilon}{2} h_{00} + 
     \frac{ \varepsilon}{2}  \sum_{k} h_{k k}, \\
\delta \left( \sqrt{-g} \langle T^{0i} \rangle \right) &= \frac{\varepsilon}{3}  h_{0 i}  ,  \\ 
\delta \left( \sqrt{-g}  \langle T^{i i} \rangle \right) &= \frac{\varepsilon}{2}h_{00} - \frac{\varepsilon}{3}  h_{i i} +
\frac{\varepsilon}{6}  \sum_{k} h_{k k}, \\
\delta \left( \sqrt{-g}  \langle T^{i j} \rangle \right) &=-\frac{\varepsilon}{3}   h_{i j} ,   \qquad i \neq j , 
\end{split}
\end{equation}
where $h_{00}=-2\sigma$, $h_{0 k}=-a_k$ and $h_{i j}= g_{i j}-\delta_{i j}$.
A look at~\eqref{static} shows that the coefficients in~\eqref{perpol} exactly agree with the components of    
$\Pi^{\mu \nu \rho \sigma}$ in the static limit.   
Thus, they encode the form of the constitutive relations for the 
energy-momentum tensor at the lowest  order in a derivative expansion of the fluid fields. 
In the remaining  of this paper we will extend Rebhan's computation to higher orders in the momenta 
and analyze the implications for the constitutive relations and parity violating partition function.


\section{Parity-odd response function}

As explained in the Introduction,  our interest will be  in  the parity violating  part of the gravitational response function.  
From the fact that  $\langle T^{\mu \nu} \rangle$ in flat space-time  is of the form  \eqref{perfect}, 
which is parity-even,  it follows that, in order to obtain the parity-odd part of 
$\delta\langle T^{\mu \nu}(x)\rangle$, eq.~\eqref{linres} can be simplified to
\begin{equation}
\label{linodd}
\delta \langle T^{\mu \nu}(x)\rangle = -\frac{1}{2} \int d^4 y\,  {\Pi}^{\mu \nu \, \rho \sigma}(x-y) 
 h_{\rho \sigma}(y) ,  
 \end{equation}
where the required response function is given by 
\begin{equation}\label{correl}
\Pi^{\mu \nu \, \rho \sigma}(x-y) \equiv -i\, \theta(x^0-y^0) \left\langle \bigl[T^{\mu \nu}(x), T^{\rho \sigma}(y)\bigr ] \right\rangle
-2\left\langle \left. \frac{\delta \bigl(\sqrt{-g(x)}T^{\mu \nu}(x)\bigr)}{\delta g_{\rho \sigma}(y)}\right|_{g=\eta} \right\rangle .  
\end{equation}

For an ideal gas of left-handed Weyl fermions, the first term in \eqref{correl} takes 
the following form in the imaginary time  formalism
\begin{equation}\label{vb1}
\begin{split}
\Pi_1^{\mu \nu \, \rho \sigma}(i \nu_n, \bm{q})&=  T \sum_{\omega_n} \int \frac{d^3 k}{(2\pi)^3} 
\text{tr} \bigl[ \mathcal{P}_- \slashed{K}V^{\mu \nu}(K, K + Q)  (\slashed{K} +  \slashed{Q} )  \\
 & \quad \times V^{\rho \sigma}(K+Q, K)    \bigr] \frac{1}{K^2 (K+Q)^2} , 
 \qquad K^0 = i \omega_n + \mu,  
\end{split}
\end{equation}
where $\mathcal{P}_- \!=\!(1 - \gamma_5)/2$ and the  fermion-fermion-graviton three-vertex, which can be read from the energy-momentum tensor $T^{\mu\nu}$ in Minkowski space~\eqref{em},  is given by\footnote{The gamma matrices obey 
$\left\{\gamma^\mu, \gamma^\nu\right\}=-2 \eta^{\mu \nu}$,
$\gamma_5 = -\tfrac{i}{24} \epsilon_{\mu \nu \rho \sigma} \gamma^\mu \gamma^\nu \gamma^\rho \gamma^\sigma$, where 
$\epsilon_{0123} = -\epsilon^{0123}=-1$. Note that we use the mostly positive signature.}
\begin{equation}\label{v3}
V^{\mu \nu}(K, P) = \frac{1}{4}\bigl[ 
 \gamma^\mu (K + P)^\nu +  \gamma^\nu (K+ P)^\mu \bigr]   
 -\frac{1}{2} \eta^{\mu \nu}  ( \slashed{K} + \slashed{P})  . 
\end{equation}
Up to parity even contributions (see appendix~\ref{seag} for details) the second term in \eqref{correl}, 
coming   from the seagull diagram, can be written
\begin{equation}\label{sea}
\begin{split}
\Pi_2^{\mu \nu \, \rho \sigma}(i \nu_n, \bm{q})&= \frac{1}{8}\eta^{\mu\rho}T\sum_{\omega_n} \int \frac{d^3 k}{(2\pi)^3} \text{tr} 
\bigl[\{\sigma^{\nu\sigma},\slashed{Q}\}\mathcal{P}_-\slashed{K}\bigr]\frac{1}{K^2}\\
&\quad +\frac{1}{8}\eta^{\nu\rho}T\sum_{\omega_n} \int \frac{d^3 k}{(2\pi)^3} \text{tr} 
\bigl[\{\sigma^{\mu\sigma},\slashed{Q}\}\mathcal{P}_-\slashed{K}\bigr]\frac{1}{K^2} +(\rho\leftrightarrow\sigma) ,
\end{split}
\end{equation}
where $\sigma^{\nu\sigma}\equiv\frac{1}{4}[\gamma^\nu,\gamma^\sigma]$. 
Here the zero subscript from $T$ and $\mu$ has been omitted. 


\subsection{Leading contribution to the parity-odd response function}

As reviewed in the previous section, for momenta $|q^0|, q \ll |\mu|, T$,  
the leading behavior of $\Pi^{\mu \nu \, \rho \sigma}$ is proportional to the energy density 
and can be written in terms the prototype integral~\eqref{four}. 
The next  to leading order contribution to $\Pi^{\mu \nu \, \rho \sigma}$ in the momentum expansion, which  is 
odd in the chemical potential and linear in $Q$,  turns out to be governed by the integral
\begin{equation}\label{i3}
I^{\mu \nu \lambda}(Q) = T \sum_{\omega_n} \int \frac{d^3 k}{(2\pi)^3} 
\frac{K^\mu K^\nu K^\lambda}{K^2 (K+Q)^2} . 
\end{equation}

Concretely, the leading parity-odd contribution to $\Pi_1$ is obtained by picking  the $\gamma_5$ 
in the projector $\mathcal{P}_- \!=\!(1 - \gamma_5)/2$ together with as many $K$s as possible in the numerator of~\eqref{vb1}. 
This  gives
\begin{equation}
\Pi^{\mu \nu \, \rho \sigma}_1 (q^0, \bm{q})= 
\frac{i}{2} Q_\alpha \eta_{\beta \lambda}\left( \epsilon^{\alpha \beta \mu \rho}
 I^{\nu \sigma \lambda} + \epsilon^{\alpha \beta \nu \rho}
 I^{\mu \sigma \lambda}\right) +(\rho\leftrightarrow\sigma)  .
\end{equation}
Similarly, the parity-odd part of $\Pi_2$ is given by 
\begin{equation}
\Pi^{\mu \nu \, \rho \sigma}_2 (q^0, \bm{q})= 
-\frac{i}{4} Q_\alpha \eta_{\beta \lambda}\left( \epsilon^{\alpha \beta \mu \rho}
 \eta^{\nu \sigma} H^\lambda + \epsilon^{\alpha \beta \nu \rho}
 \eta^{\mu \sigma} H^\lambda \right) +(\rho\leftrightarrow\sigma) ,
\end{equation}
where
\begin{equation}
 H^\mu=T \sum_{\omega_n} \int \frac{d^3 k}{(2\pi)^3}  
\frac{K^\mu }{K^2 } . 
\end{equation}

As before, the evaluation of~\eqref{i3} involves Matsubara summation and  rescaling of the resulting integrand through   
$q^0 \to \epsilon q^0, \bm{q} \to \epsilon \bm{q}$.  
The integral  obtained by keeping the $\mathcal{O}(\epsilon^0)$ term, which we will denote $I^{\mu \nu \lambda}_{(0)}$,   
has all  components proportional to 
\begin{equation}
I_{(0)}= \int_0^\infty \frac{dk}{4\pi^2} k^2 \bigl(n_F(k-\mu) - n_F(k+\mu)\bigr) = \frac{1}{12\pi^2} (\mu^3 + \pi^2 \mu\, T^2) . 
\end{equation}
In particular, we have
\begin{equation}
\eta_{\alpha \beta} I^{\alpha \beta \gamma}_{(0)}  = H^\gamma =-I_{(0)} u^\gamma, 
\end{equation} 
where $u^\gamma = (1,\bm{0})$ is the velocity of the fluid in the local rest frame.  
Using this relation to eliminate $H^\lambda$ in favor of $I^{\mu \nu \lambda}_{(0)}$ and adding the contributions of $\Pi_1$ and $\Pi_2$,  gives the following expression for the leading parity-odd contribution to the polarization tensor
\begin{equation}\label{pi1}
\begin{split}
\Pi^{\mu \nu \, \rho \sigma}_{(1)} (q^0, \bm{q})&= 
\frac{i}{2} Q_\alpha \eta_{\beta \lambda} \epsilon^{\alpha \beta \mu \rho}
 \Bigl(I_{(0)}^{\nu \sigma \lambda} - \frac{1}{2} \eta^{\nu \sigma} \eta_{\gamma \kappa} I_{(0)}^{\gamma \kappa \lambda} \Bigr) +\ldots  \\
&=\frac{i}{4} Q_\alpha \eta_{\beta \lambda} 
I_{(0)}^{\gamma \kappa \lambda} 
  \bigl[
 \epsilon^{\alpha \beta \mu \rho} \bigl(\delta_{\gamma \kappa}^{\nu \sigma} - 
                  \eta^{\nu \sigma} \eta_{\gamma \kappa}\bigr)  \\ 
                  & \quad +    
  \epsilon^{\alpha \beta \nu \rho} \bigl(\delta_{\gamma \kappa}^{\mu \sigma} - \eta^{\mu \sigma} \eta_{\gamma \kappa}\bigr) 
+( \rho \leftrightarrow \sigma) \bigr] ,    
\end{split}
\end{equation}
where $\delta_{\gamma \kappa}^{\nu \sigma} = \delta_\gamma^\nu \delta_\kappa^\sigma + 
\delta_\gamma^\sigma \delta_\kappa^\nu$.

We can check this result by noting that the covariant conservation law for the energy-momentum tensor 
imposes constraints on the tensorial structure of $\Pi^{\mu \nu \, \rho \sigma} (Q)$. 
Since the covariant derivative involves  a combination of derivatives of 
$\delta \langle T^{\mu \nu} \rangle$ and  Christoffel symbols,  
 and both of them  are of the same order 
in $h_{\rho \sigma}$, it follows that the vanishing of  $\nabla_\mu \delta \langle T^{\mu \nu}(x)\rangle$ gives rise to Ward identities connecting  $Q_\mu \Pi^{\mu \nu \, \rho \sigma} (Q)$ with 
the one-point function $\langle T^{\alpha \beta}\rangle$~\cite{Freedman:1974xm}.
The consequences of this requirement in relation with~\eqref{four} were fully analyzed in~\cite{Rebhan:1990yr}.  
With regard to the parity-violating part  of $\Pi^{\mu \nu \, \rho \sigma} (Q)$, the conservation law imposes uniquely 
$Q_\mu \Pi^{\mu \nu \, \rho \sigma} (Q)=0$,  since the parity-violating part of $\langle T^{\alpha \beta}\rangle$ vanishes. 
Similarly, conformal invariance requires  $\eta_{\mu \nu}\Pi^{\mu \nu \, \rho \sigma} (Q)=0$ for this part.
Now, $\Pi_{(1)}^{\mu \nu \, \rho \sigma}$ as given by~\eqref{pi1} is obviously traceless, while transversality follows from the following property satisfied by the componets of $I^{\mu \nu \lambda}$  at leading order
\begin{equation}
\label{trQ}
I_{(0)}^{\mu \nu \lambda}  Q_\lambda =   \frac{1}{2} (Q^\mu \delta^\nu_\gamma + 
 Q^\nu \delta^\mu_\gamma) \eta_{\alpha \beta} I_{(0)}^{\alpha \beta \gamma} ,   
\end{equation} 
which is easily proved from the results in table~\ref{t1} of appendix~\ref{appi3}. 
Actually, up to an overall normalization, the structure of  the leading parity-odd response function~\eqref{pi1} is uniquely determined by the conditions of transversality and tracelessness together with~\eqref{trQ}.

In order to evaluate $\Pi^{\mu \nu \, \rho \sigma}_{(1)}$ we need explicit expressions for all the components of $I^{\mu \nu \lambda}$. This is conveniently done by 
using the decomposition $I^{\mu \nu \lambda} = \sum_j c_j T^{\mu \nu \lambda}_j$
in terms of the six symmetric tensors 
\begin{equation}\label{tensors6}
\begin{split}
T^{\mu \nu \lambda}_1 &= Q^\mu Q^\nu Q^\lambda, \\ 
T^{\mu \nu \lambda}_2 &= Q^\mu Q^\nu u^\lambda+Q^\mu u^\nu Q^\lambda+u^\mu Q^\nu Q^\lambda, \\ 
T^{\mu \nu \lambda}_3 &= Q^\mu u^\nu u^\lambda+u^\mu Q^\nu u^\lambda+u^\mu u^\nu Q^\lambda , \\ 
T^{\mu \nu \lambda}_4 &= Q^\mu \eta^{\nu \lambda} + Q^\nu \eta^{\mu \lambda}+Q^\lambda \eta^{\mu \nu} , \\ 
T^{\mu \nu \lambda}_5 &= u^\mu u^\nu u^\lambda, \\ 
T^{\mu \nu \lambda}_6 &= u^\mu \eta^{\nu \lambda} + u^\nu \eta^{\mu \lambda}+u^\lambda \eta^{\mu \nu} .
\end{split}
\end{equation}
The contractions of $I^{\mu \nu \lambda}$ with the basis tensors are given in table~\ref{t1} of appendix~\ref{appi3}. 
Inverting these relations gives the coefficients $c_j$ 
\begin{equation}\label{i30}
\begin{split}
c_1(q^0,q)/I_{(0)} &= 
     -\frac{5 q^0}{4 q^4} - \frac{3 q^0(5 Q^2-2q^2)}{4 q^6} L(q^0, q), \\ 
c_2(q^0,q)/I_{(0)} &=
 -\frac{5 Q^2-2 q^2}{4 q^4}  - 
                                             \frac{3Q^2(5Q^2-4q^2)}{4 q^6} L(q^0, q) ,  \\ 
c_3(q^0,q)/I_{(0)} &=    
 \frac{q^0\left(5 Q^2 +2 q^2\right)}{4 q^4}  +
                                            \frac{15 q^0 Q^4}{4 q^6}   L(q^0, q) , \\ 
c_4(q^0,q)/I_{(0)} &=  
\frac{q^0}{4 q^2} + \frac{3 q^0Q^2}{4 q^4} L(q^0, q),  \\ 
c_5(q^0,q)/I_{(0)} &= 
Q^2\left(\frac{5 Q^2 +2 q^2}{4 q^4} +
                                           \frac{15  Q^4}{4 q^6}   L(q^0, q)  \right) , \\    
c_6(q^0,q)/I_{(0)} &=  Q^2\left(\frac{1}{4 q^2} +\frac{3 Q^2}{4 q^4}  L(q^0, q)  \right)  ,  
\end{split}                                                              
\end{equation}
where  $ L(q^0, q)\equiv Q_1(q^0/q)$ is the Legendre function of the second kind,  which results from using the retarded prescription 
$ i \nu_n + 0^+  \to q^0$ in the  analytic continuation   of the integral
\begin{equation}\label{lq0}
\int_{-1}^1dt \frac{t}{i \nu_n - q t}  \to 
\frac{2}{q} L(q^0, q) =\frac{2}{q}\left[-1+
 \frac{q^0}{2 q} \ln\biggl|\frac{q^0+q}{q^0-q} \biggr| - \frac{i \pi}{2}\frac{q^0}{q}  \theta\Bigl(1 - \frac{(q^0)^2}{q^2} \Bigr)
 \right]. 
\end{equation}    

This result can be greatly simplified  by  using the 
constraints imposed by \eqref{trQ},  which restrict the number of independent functions to three, 
for example $c_1, c_2$ and $c_3$. Clearly $c_1$ does not contribute, and  
we are left with only two combinations of the functions $c_2$ and $c_3$.  
Due to the transversality of the response function, the result can be written in a 
particularly transparent way in terms of  the two projectors $P_\mathbb{T}$ and $P_\mathbb{V}$, 
\begin{equation}
\begin{split}
 P_\mathbb{T} ^{\mu \nu} &= \eta^{\mu \nu} - \frac{1}{(u \cdot Q)^2 + Q^2} \left[  
     u \cdot Q  \bigl(u^\mu Q^\nu + u^\nu Q^\nu ) + Q^\mu Q^\nu-Q^2u^\mu u^\nu \right] , \\ 
  P_\mathbb{V}^{\mu \nu} &= \eta^{\mu \nu}- \frac{Q^\mu Q^\nu}{Q^2}  - P_\mathbb{T}^{\mu \nu}  .
 \end{split}
\end{equation}
Then the response function \eqref{pi1} adopts the simple form
\begin{equation}\label{proy} 
\begin{split}
\Pi_{(1)}^{\mu \nu \, \rho \sigma}(q^0, \bm{q})&=
i c_{\mathbb{V}}(q^0, q) \frac{Q^2}{(u \cdot Q)^2 + Q^2} u_\alpha Q_\beta \bigl[ \epsilon^{\alpha \beta \mu \rho}
    P_\mathbb{V} ^{\nu \sigma}  + 
    \epsilon^{\alpha \beta \nu \rho}
    P_\mathbb{V} ^{\mu \sigma}  +   ( \rho \leftrightarrow \sigma ) \bigr]    \\ 
    & \quad   + 
i c_{\mathbb{T}}(q^0, q) u_\alpha Q_\beta \bigl[ \epsilon^{\alpha \beta \mu \rho}
    P_\mathbb{T} ^{\nu \sigma}  + 
    \epsilon^{\alpha \beta \nu \rho}
    P_\mathbb{T} ^{\mu \sigma}  +   ( \rho \leftrightarrow \sigma ) \bigr] ,  
\end{split} 
\end{equation}
where the functions $c_{\mathbb{L,T}}$ are given by  
\begin{equation}\label{primer} 
\begin{split}
c_\mathbb{V}(q^0, q)&= \frac{q^2}{2 q^0}(q^0 c_2 + c_3)  \\ 
  &=  \frac{1}{24\pi^2} \left(\mu^3 + \pi^2 \mu  T^2 \right)  
 \left( 1  + \frac{3Q^2}{q^2} L(q^0, q) \right) , \\   
c_\mathbb{T}(q^0, q)&= \frac{1}{2}\left[Q^2 c_2 - q^0 c_3 \right]    \\ 
   &=-\frac{1}{96\pi^2}\left(\mu^3 + \pi^2  \mu T^2 \right) 
  \left(2 + \frac{Q^2}{q^2}  + \frac{3Q^4}{q^4} L(q^0, q) \right) . 
\end{split}
\end{equation}
In the static limit these become
\begin{equation}\label{c1static}
\begin{split}
c_{\mathbb{V}}(0,q) &= -\frac{1}{12 \pi^2} \left(\mu^3 + \pi^2 \mu \, T^2 \right), \\  
c_{\mathbb{T}}(0,q) &=  0.
\end{split}
\end{equation}
These are the main results in this subsection. 

\subsection{Parity-odd response function at higher orders  in the momenta}
In order to use  the response function as a source of 
constraints on the anomalous constitutive relations at higher order in the derivative expansion, 
we will need   ${\Pi}^{\mu \nu \rho \sigma}(Q)$ up to third order in $Q$. 
The form of the second order contribution comes from the explicit contribution of order $Q^2$ to  
the trace in \eqref{vb1} together with the piece of  $\mathcal{O}(Q)$ in  $I^{\alpha \beta \gamma}$.  
From the explicit formula~\eqref{sea} for $\Pi_2$, it is obvious that the seagull diagram does not contribute beyond  first order. 
Therefore the second order correction reads 
\begin{equation}\label{pi2}
\begin{split}
\Pi^{\mu \nu \, \rho \sigma}_{(2)} (Q) =  &\frac{i}{2} Q_\alpha \eta_{\beta \lambda} \epsilon^{\alpha \beta \mu \rho}
 \Bigl(I_{(1)}^{\nu \sigma \lambda} + \frac{1}{2} \bigl(Q^\nu I_{(0)}^{\sigma \lambda} + Q^\sigma I_{(0)}^{\nu \lambda}  \bigr) \Bigr)  \\
 & \frac{i}{2} Q_\alpha \eta_{\beta \lambda} \epsilon^{\alpha \beta \nu \rho}
 \Bigl(I_{(1)}^{\mu \sigma \lambda} + \frac{1}{2} \bigl(Q^\mu I_{(0)}^{\sigma \lambda} + Q^\sigma I_{(0)}^{\mu \lambda}  \bigr) \Bigr) +(\rho\leftrightarrow\sigma) ,
 \end{split} 
\end{equation}
where we have introduced the integral 
\begin{equation}\label{second}
I^{\alpha \beta}(Q) = T \sum_{k^0} \int \frac{d^3 k}{(2\pi)^3} 
\frac{K^\alpha K^\beta}{K^2 (K+Q)^2} ,
\end{equation}
which in this case  is needed only at $\mathcal{O}(Q^0)$. A computation using  the  results for the integrals in appendices~\ref{appi3} and~\ref{appi2} shows that, although neither $I_{(0)}^{\alpha \beta}$  nor 
$I_{(1)}^{\alpha \beta \gamma}$ vanish, the  specific combination in~\eqref{pi2} does. 
Therefore the second order contribution to the parity odd gravitational response function vanishes identically. 

The third order contribution may be expressed similarly. Besides   
$I^{\alpha \beta \gamma}$ at $\mathcal{O}(Q^2)$ and    $I^{\alpha \beta}(Q)$ at $\mathcal{O}(Q)$, we also need the $\mathcal{O}(Q^0)$ piece of the integral $I^{\mu}$, which 
turns out to be independent of the temperature
\begin{equation}
\label{oneI}
I^{\mu} = T \sum_{k^0} \int \frac{d^3 k}{(2\pi)^3} 
\frac{K^\mu}{K^2 (K+Q)^2} \sim \frac{\mu L(q^0, q)}{8 \pi^2}\left(\frac{Q^2}{q^2}  u^\mu + 
\frac{q^0 }{q^2}  Q^\mu \right) . 
\end{equation}
The appropriate combination is now  
\begin{equation}\label{third}
\begin{split}
\Pi_{(3)}^{\mu \nu \, \rho \sigma} (q^0, \bm{q}) &= 
\frac{i}{2} Q_\alpha \eta_{\beta \lambda} \epsilon^{\alpha \beta \mu \rho} 
\Bigl( I_{(2)}^{\nu \sigma \lambda} +  \frac{1}{2} \bigl(Q^\nu I_{(1)}^{\sigma \lambda} + Q^\sigma I_{(1)}^{\nu \lambda}  \bigr)
+ \frac{1}{4} Q^\nu Q^\sigma I_{(0)}^\lambda \Bigr) \\ 
& \quad + \frac{i}{2} Q_\alpha \eta_{\beta \lambda} \epsilon^{\alpha \beta \nu \rho} 
\Bigl( I_{(2)}^{\mu \sigma \lambda} +  \frac{1}{2} \bigl(Q^\mu I_{(1)}^{\sigma \lambda} + Q^\sigma I_{(1)}^{\mu \lambda}  \bigr)
+\frac{1}{4} Q^\mu Q^\sigma I_{(0)}^\lambda \Bigr) + (\rho \leftrightarrow \sigma ) . 
 \end{split}
\end{equation}
Using  the explicit results for the integrals in appendices~\ref{appi3} and~\ref{appi2}, and following the method outlined in the previous subsection, we find that  
the third order response function is still given by   eq.~\eqref{proy}, with  the coefficients replaced by
\begin{equation}\label{tercer}
\begin{split}
c_{\mathbb{V}} &= \frac{\mu\,  q^2}{192 \pi^2} \left[ -\frac{2 Q^2}{q^2}   + 
 \frac{3 Q^2 (q^2-2 Q^2)}{q^4}  L(q^0, q) \right] , \\  
c_{\mathbb{T}} &=  \frac{\mu \, q^2}{192 \pi^2}\left[ \frac{ Q^4}{2 q^4}  + 
      \frac{3Q^6}{2 q^6}  L(q^0, q) \right].
\end{split}
\end{equation}
In the static limit they become 
\begin{equation}
\begin{split}\label{c3static}
c_{\mathbb{V}}(0,q) &= \frac{\mu}{192 \pi^2} q^2, \\  
c_{\mathbb{T}}(0,q) &=  -\frac{\mu}{192 \pi^2} q^2.
\end{split}
\end{equation}
This completes the computation of the parity-odd response function to third order in $Q^\nu$. 

 
\section{Energy-momentum tensor and metric perturbations}

In this section  we will use the results  obtained for the parity-odd response function  to derive the 
general form of the parity violating part of the energy-momentum tensor. 
We will devote special attention  to the static case, where our results can be compared with recent proposals in the literature.
As shown above, the effects of metric perturbations on the energy-momentum tensor 
\begin{equation}
\label{lrq}
 \delta\langle T^{\mu\nu}\rangle=-\frac{1}{2} \Pi^{\mu \nu \, \rho \sigma} (q^0, \bm{q}) h_{\rho \sigma} ,
\end{equation}
can be parametrized by  the two independent functions $c_{\mathbb{V}}(q^0, q)$ and $c_{\mathbb{T}}(q^0, q)$ 
\begin{equation}
\begin{split}
\Pi^{\mu \nu \, \rho \sigma}(q^0, \bm{q})&=
i c_{\mathbb{V}}(q^0, q) \frac{Q^2}{(u \cdot Q)^2 + Q^2} u_\alpha Q_\beta \bigl[ \epsilon^{\alpha \beta \mu \rho}
    P_{\mathbb{V}} ^{\nu \sigma}  + 
    \epsilon^{\alpha \beta \nu \rho}
    P_{\mathbb{V}} ^{\mu \sigma}  +   ( \rho \leftrightarrow \sigma ) \bigr]    \\ 
    & \quad   + 
i c_{\mathbb{T}}(q^0, q) u_\alpha Q_\beta \bigl[ \epsilon^{\alpha \beta \mu \rho}
    P_{\mathbb{T}} ^{\nu \sigma}  + 
    \epsilon^{\alpha \beta \nu \rho}
    P_{\mathbb{T}} ^{\mu \sigma}  +   ( \rho \leftrightarrow \sigma ) \bigr]  .
\end{split} 
\end{equation}

\begin{table}[t!]
\begin{center}
\begin{tabular}{r || c |c| c  }
  & Scalar & Vector & Tensor \\
\hline \hline
$h_{00}$ & 
$-2 \sigma$ & 
-- &
-- \\
$h_{0i}$&
$-\partial_i b$ &
$-a_i^{(S)}$ &
-- \\
$h_{ij}$ &
$c\delta_{ij}+\partial_i\partial_j d$ &
$\partial_i F_j+\partial_j F_i$ &
$\tilde h_{ij}$\\
 \hline
\end{tabular}
\caption{$SO(3)$ components of a general  perturbation of the metric, 
where we have defined $a_i^{(S)}$ and  $a_i^{(L)}=\partial_i b$ as the solenoidal and irrotational parts of $\delta g_{0i}=-a_i(t, \bm{x})$.}
\label{t3}
\end{center}
\end{table}
 
It should be noted that, as a consequence of the form of $\Pi^{\mu \nu \, \rho \sigma}$, 
the response of the energy-momentum tensor depends only on gauge-invariant combinations of the metric disturbances. 
In order to simplify the analysis, a general metric perturbation has been decomposed into $SO(3)$ irreducible components
in table~\ref{t3}. Note that the vector fields  $a^{(S)}_i$ and $F_{i}$ are solenoidal, 
while $\tilde h_{ik}$ is traceless and satisfies $\partial_i\tilde h_{ik}=0$.
Direct substitution shows that scalar perturbations  do not produce any parity-violating effect  on $T^{\mu \nu}$. 
For vector perturbations, not necessarily time-independent,  
the change in  the expectation value of $\langle T^{\mu \nu}\rangle$ 
depends only on the combination $a_i  +  \partial_t F_i$, which is gauge-invariant 
 \begin{equation}
 \label{pvector}
 \begin{split}
 \delta\langle T^{0 i}\rangle&= c_\mathbb{V}(q^0, q)  \, i \epsilon^{i j k}  q^j (-a_k + i q^0 F_k), \\ 
 \delta\langle T^{i j}\rangle&= c_\mathbb{V}(q^0, q) \, i q^0 
 \left( \epsilon^{i m n}  \hat{q}^m \hat{q}^j + \epsilon^{j m n}  \hat{q}^m \hat{q}^i  \right)(-a_n + i q^0 F_n) ,
 \end{split}
 \end{equation}
 where $\hat{q}^j=q^j/q$. 
 Thus, $c_\mathbb{V}(q^0, q)$ parametrizes the response to vector perturbations of the metric. 
 Similarly, $c_\mathbb{T}(q^0, q)$ parametrizes the response to tensor perturbations $\tilde h_{ij}$, 
 which are  gauge-invariant by construction
 \begin{equation}
 \label{ptensor}
 \delta\langle T^{i j}\rangle= -c_\mathbb{T}(q^0, q)\epsilon^{ilm} \delta^{j n}\, i q^l \tilde h_{m n}
 +   ( i \leftrightarrow j ) .
 \end{equation}
These are the main results in this paper. In what follows, we will compare them with other results in the literature.


\subsection{Static limit of response functions and anomalous constitutive relations}
Arguments based on linearized   
hydrodynamics~\cite{Kadanoff} show that  the small velocity field of the fluid
$v^j(t, \bm{x})$ is the 
quantity that plays the role of external  force coupled to the momentum density $T^{0 i}$ in the perturbing Hamiltonian 
\begin{equation}
 H^\text{ext} = -\int d^3 \bm{x}\, T^{0i} (t,\bm{x}) v^i(t, \bm{x}).   
 \end{equation}
Hence a comparison with 
\begin{equation}
T^{\mu \nu} = \frac{2}{\sqrt{-g}} \frac{\delta S}{\delta g_{\mu \nu}}, 
\end{equation}   
enables us  to identify $h_{0i} =-a_i$ 
with the fluid velocity $v^i$ in the static limit $q^0=0$. 
Thus, using~\eqref{c1static}  we find the following constitutive relations  in the static limit  at leading order in the momenta
\begin{equation}
\label{eq:constpv}
\begin{split}
\delta\langle T^{0 i}\rangle &= c_{\mathbb{V}}(0, q) \epsilon^{i j k} i q^j v^k  , \\ 
\delta\langle T^{ ij}\rangle &=0 .
\end{split} 
\end{equation}
We see that $c_{\mathbb{V}}(0, q)$ may be interpreted as a parity-violating susceptibility  connecting 
the vorticity with the momentum density. In fact, in hydrodynamics with quantum anomalies,  
 the momentum density  to linear order in $\bm{v}$~\cite{Loganayagam:2011mu}  is precisely given by
\begin{equation}\label{md}
\bm{g} =(\varepsilon + \mathcal{P}) \bm{v} + \chi_{\mathbb{V}}  \bm{\nabla}\times \bm{v} , 
\end{equation}
when one uses a frame where the entropy current does not have anomalous part i.e., where 
$J_S^\mu = s u^\mu$.
This agrees with our result~\eqref{eq:constpv} which, by~\eqref{c1static}, implies the following value for the anomalous susceptibility
\begin{equation}\label{as}
 \chi_{\mathbb{V}} =c_{\mathbb{V}}(0, q)=-\frac{1}{12 \pi^2} \left(\mu^3 + \pi^2 \mu T^2 \right) . 
\end{equation}
Actually, the value of the parity violating susceptibility has been  related to the gauge and 
mixed anomaly coefficients in the anomalous conservation equations
\begin{align}
\nabla_\mu J_{\mathrm{cov}}^{\mu} &= \frac{1}{4} \epsilon^{\mu \nu \rho\sigma} \bigl(
3 c_A  F_{\mu \nu} F_{\rho\sigma}  +c_m 
 R^{\alpha} {}_{\beta\mu\nu}   R^{\beta}{}_{\alpha \rho\sigma}  \bigr) ,\\
 \nabla_\nu T_{\mathrm{cov}}^{\mu\nu} &= F^\mu{}_{\nu} J_{\mathrm{cov}}^{\nu}+ 
     \frac{1}{2}c_m\nabla_\nu \bigl(\epsilon^{ \rho\sigma\alpha\beta}  F_{\rho\sigma} 
     R^{\mu\nu}{} _{\alpha\beta}   \bigr) ,
\end{align}
through the relations~\cite{Jensen:2012kj}
\begin{equation}
 \chi_{\mathbb{V}}=2 (\tilde c_{4d}\,\mu\, T^2-c_A \mu^3) , \qquad \tilde c_{4d}=-8 \pi^2 c_m .
\end{equation}
The values of the anomaly coefficients for a left-handed spinor in $(3+1)$ dimensions,
\begin{equation}
c_A=8 c_m= \frac {1}{24 \pi^2} ,
\end{equation}
then imply our value~\eqref{as} for the anomalous susceptibility. 

Similarly, we may use~\eqref{c3static} to find the following $\mathcal{O}(Q^3)$ corrections 
to the constitutive relations in the static limit 
\begin{align}
\label{vecok}
\delta\langle T^{0 i}\rangle &= \frac{\mu}{192\pi^2}\epsilon^{i j k} \nabla^2\partial_j a_k ,  \\   
\label{tenok}
\delta\langle T^{ ij}\rangle &= -\frac{\mu}{192\pi^2}\epsilon^{i l m} \delta^{jk}\nabla^2\partial_l\tilde h_{km}+
 (i \leftrightarrow j) .
 \end{align} 
Note that \eqref{vecok} simply gives a correction to the momentum density~\eqref{md}
\begin{equation}
\delta\bm{g} = -\frac{\mu}{192\pi^2}\nabla^2( \bm{\nabla}\times \bm{v})=-c_m\mu \nabla^2( \bm{\nabla}\times \bm{v}) .
\end{equation}
On the other hand, \eqref{tenok} describes a qualitatively new effect. For a tensorial perturbation depending only on $z$ 
\begin{equation}
\label{tensor4}
\tilde h_{i j}=
\begin{pmatrix}
h_+(z) & h_\times(z) & 0 \\ 
h_\times(z) & -h_+(z) & 0 \\ 
0 & 0 &0
\end{pmatrix} , 
\end{equation}
eq.~\eqref{tenok} gives 
 \begin{equation}
 \begin{split}\label{tensorz}
  T^{11} &=  -T^{22} = 2 c_m\mu  h_{\times}'''(z) ,  \\ 
 T^{12} &=  -2 c_m\mu  h_{+}'''(z)  . 
 \end{split}
\end{equation}

Predictions for the effects of gauge and mixed anomalies on the constitutive relations at higher orders order in the momenta have only recently been given in $(1\!+\!1)$~\cite{Valle:2012em} and $(3\!+\!1)$ dimensions \cite{Jensen:2012kj}. In order to connect our results with those in~\cite{Jensen:2012kj} we collect some of their formulae.  
Using a  frame where the energy-momentum tensor takes the form
\begin{equation}
\langle T^{\mu \nu}\rangle =  (\varepsilon +P) u^\mu u^\nu + P g^{\mu \nu}+u^\mu q_A^\nu+
u^\nu q_A^\mu+\tau_A^{\mu \nu} ,
\end{equation}
with $u_\mu q_A^\mu=u_\mu \tau_A^{\mu \nu}=\tau^\mu_{A\;\mu}=0$, 
the following results were obtained~\cite{Jensen:2012kj} at $\mathcal{O}(Q^3)$ in  $(3\!+\!1)$ dimensions
\begin{align}
q_A^\mu&=-2\mu c_m \tilde v_3^\mu=-2\mu c_m  \epsilon^{\mu\nu\rho\sigma}u_\nu \nabla_\rho R_{\sigma\alpha}u^\alpha+\ldots \\
\tau_A^{\mu \nu}&= 4\mu c_m \tilde t_3^{\mu \nu}=4\mu c_m\Delta^{\alpha <\mu}\epsilon^{\nu>\rho\sigma\beta} u_\rho\nabla_\sigma R_{\alpha\beta}+\ldots
\end{align}
where the dots stand for contributions that vanish at linear order in the metric perturbations and are thus invisible in our computation. In this formula
\begin{equation}
V^{<\mu\nu>}\equiv \Delta^{\mu\rho} \Delta^{\nu\sigma} V_{(\mu\nu)}-\frac{1}{3} \Delta^{\mu\nu}\Delta_{\rho\sigma}V^{\rho\sigma}
\end{equation} 
and $ \Delta^{\mu\nu}=g^{\mu\nu}+u^\mu u^\nu$. In order to compare with~\eqref{vecok}, we evaluate the pseudovector $\tilde v_3^\mu$ at linear order in the metric perturbation. This yields $\tilde v_3^0\!=\!0$ and
\begin{equation}
\tilde v_3^i=-\frac{1}{2}\epsilon^{ijk}\left(\partial_j\nabla^2 h_{k0}- \partial_0\partial_j\partial_lh_{lk}\right)+\mathcal{O}(h^2) , 
\end{equation}
where we have taken $u^\mu=\delta^\mu_0+\mathcal{O}(h)$.  Upon substitution in this expression of the metric components in table~\ref{t3} we obtain
\begin{equation}
q_A^i=-2\mu c_m \tilde v_3^i=-\frac{\mu}{192\pi^2}\epsilon^{i j k} \nabla^2\partial_j(a_k+\partial_t F_k)+\mathcal{O}(h^2) , 
\end{equation}
which differs in sign from our contribution~\eqref{vecok} to $\delta\langle T^{0 i}\rangle$ in the static limit. 
On the other hand, note  that precisely the combination $a_k+\partial_t F_k$, which we have argued should be identified with the fluid velocity, appears in this expression. Proceeding similarly with the pseudotensor gives the non-vanishing components
\begin{equation}
\tilde t_3^{ij}=-\frac{1}{4} \epsilon^{ilm}\left(\partial_l\partial^\alpha\partial_\alpha h_m{}^j - 
\partial_l\partial^j\partial_\alpha h_m{}^\alpha \right)+( i \leftrightarrow j )+\mathcal{O}(h^2) .
\end{equation}
Taking the static limit and substituting the metric components in table~\ref{t3} gives
\begin{equation}
\tau_A^{ij}=4\mu c_m \tilde t_3^{ij}=
-\frac{\mu}{192\pi^2}\epsilon^{i l m}\nabla^2\partial_l\tilde h_m{}^j+( i \leftrightarrow j ) ,
\end{equation}
which exactly agrees with~\eqref{tenok}. In the next section we will explore the relations of our linearized results with the full anomalous partition function given in~\cite{Jensen:2012kj} and will be able to explain the sign discrepancy noted above.


\section{The anomalous partition function and the static response function}  

In this section we consider in more detail  how  our previous results for the  static response functions fit  
with known facts about gauge and gravitational anomalies of the underlying theory, 
particularly with the form of 
the full anomalous partition function. As we will see, 
an important feature of the third order response \eqref{c3static}  is 
that it reveals the existence of an invariant  contribution to the partition function 
which, in principle,  does not seem to arise from general arguments about anomalies.

We will assume that an external time-independent gauge field $\mathcal{A}=\mathcal{A}_0(\bm{x})dt + 
\mathcal{A}_i(\bm{x}) dx^i$ is present besides  the static gravitational field given in \eqref{backg}.
In this background  with  Killing vector $V^\mu=(1,\bm{0})$, 
the equilibrium temperature and  chemical potential become position-dependent. 
They are defined in terms of the invariant length of the time circle  and the Polyakov loop $P_A$ as~\cite{Jensen:2012jh} 
\begin{equation}
\begin{split}
T(\bm{x})^{-1} &= \int_0^{1/T_0} d\tau\sqrt{-V^\mu V_\mu} = \frac{e^\sigma}{T_0},  \\ 
 \frac{\mu(\bm{x})}{T(\bm{x})} &= \ln P_A =  \int_0^{1/T_0} d\tau \mathcal{A}_\mu(\bm{x}) V^\mu =  
 \frac{ \mathcal{A}_0(\bm{x})}{T_0},   
\end{split}
\end{equation}  
where $T_0^{-1}$  is the length of the compactified  imaginary time.

\subsection{The first order generating functional} 

Let us start with the first order in the derivative expansion. 
Using the definition of the energy-momentum tensor in terms of the partition function\footnote{Here 
$W =\ln Z= -\Omega/T_0$, where $Z$ is the grand partition function and $\Omega$ refers to the thermodynamic potential, 
a functional of the background quantities depending solely on $\bm{x}$.}
\begin{equation}
\label{currents}
T^{\mu \nu}  = \frac{2 T_0}{\sqrt{-g}} \frac{\delta W}{\delta g_{\mu \nu}(\bm{x})} ,   
\end{equation}
the linearized constitutive relation~\eqref{eq:constpv}  may be rewritten
in terms of $\mathcal{A}_0$ and $T_0$ (or equivalently $\mu$ and $T$)
\begin{equation}\label{equiv}
\delta \langle T^{0 i} \rangle = -T_0 \frac{\delta W}{\delta a_i} 
= 
(2 c_A \mathcal{A}_0^3 - 2 \tilde c_{4d} T_0^2 \mathcal{A}_0) \widetilde{\epsilon}^{i j k} \partial_j a_k + \ldots , 
\end{equation}
where $\tilde{\epsilon}^{123} = 1$, and $\epsilon^{i j k} = \tilde{\epsilon}^{i j k}/\sqrt{g_3}$  
will denote the corresponding tensor.  
Since the quantities $a_j$ and $\mathcal{A}_i$ transform like the components of a
covariant vector under spatial diffeomorphisms  
and $\mathcal{A}_0$ behaves like a scalar,  it is clear that $W$ must include the terms
\begin{equation}
 -\frac{c_A}{T_0} \int d^3 x \sqrt{g_3}\, \epsilon^{i j k} \mathcal{A}_0^3 \,  a_i  \partial_j a_k + 
  \tilde c_{4d} T_0 \int d^3 x \sqrt{g_3} \, \epsilon^{i j k} \mathcal{A}_0 \,  a_i  \partial_j a_k . 
 \end{equation} 
To reconstruct the full dependence of $W$ on the gauge field, 
we complete the action with the requirements of gauge invariance up to a $U(1)$ anomaly, 
and invariance under Kaluza-Klein transformations~\cite{Banerjee:2012iz}. 
These correspond to redefinitions of time, $t \to t' =t+ \phi(\bm{x})$, without change in the spatial coordinate, 
and preserve the form of the metric if $a_i$ transforms as $\delta a_i = -\partial_i \phi$.  
Under such a transformation the gauge field changes as $\delta \mathcal{A}_0 = 0$, 
$\delta \mathcal{A}_i = -\mathcal{L}_\phi \mathcal{A}_i =  -\mathcal{A}_0  \partial_i \phi$, and  the 
combination  $\mathcal{A}_i - a_i \mathcal{A}_0$ remains invariant. 
As this kind of gauge invariance may be viewed as a manifestation of the underlying  diffeomorphism invariance, 
which is not anomalous at the first  derivative order, it is natural to impose  this requirement on the partition function.   
A short computation shows that, up to a total derivative, 
the resulting Kaluza-Klein invariant action is given by
\begin{equation}\label{anom1}
\begin{split}
W &= -\frac{c_A}{T_0} \int d^3 x \sqrt{g_3}\, \epsilon^{i j k} \mathcal{A}_0^3 \,  a_i  \partial_j a_k 
          +\frac{3 c_A}{T_0} \int d^3 \sqrt{g_3} x\, \epsilon^{i j k} \mathcal{A}_0^2  \mathcal{A}_i  \partial_j a_k \\ 
      &\quad - \frac{2 c_A}{T_0} \int d^3 x \sqrt{g_3}\, \epsilon^{i j k} \mathcal{A}_0 \mathcal{A}_i  \partial_j \mathcal{A}_k  - 
      \tilde c_{4d} T_0  \int d^3 x \sqrt{g_3}\, \epsilon^{i j k}(\mathcal{A}_i- \mathcal{A}_0 \,  a_i ) \partial_j a_k . 
\end{split}
\end{equation}
Thus  the static gravitational response, together with the requirements  of invariance under Kaluza-Klein transformations and  gauge invariance up to a $U(1)$ anomaly,  determine the  partition function at first order in the derivative expansion. 
The first three terms in~\eqref{anom1} constitute the anomalous part of the consistent partition function, while the last one is gauge invariant. Our result~\eqref{anom1} agrees with the form of the parity-odd partition function 
proposed in~\cite{Banerjee:2012iz} for a CPT invariant theory.
The consistent gauge anomaly follows from the variation of this action under a gauge transformation 
$\delta \mathcal{A}_i = \partial_i \Lambda$, $\delta \mathcal{A}_0 = 0$. This induces a change 
\begin{equation}
\delta W = \frac{2 c_A}{T_0} \int d^3 x \sqrt{g_3}\, \Lambda \epsilon^{i j k}
  \partial_i \mathcal{A}_0 \partial_j \mathcal{A}_k , 
\end{equation} 
which shows that the consistent anomaly is precisely determined by the cubic term  in the static momentum density correlator. 
The variation of the consistent current under such a gauge transformation is given by 
\begin{equation}
\begin{split}
\delta J^0 &= -2 c_A \frac{\tilde{\epsilon}^{i j k}}{\sqrt{-g}}  \partial_j \mathcal{A}_k \partial_i \Lambda,  \\ 
\delta J^i &= -2 c_A \frac{\tilde{\epsilon}^{i j k}}{\sqrt{-g}}  \partial_j \mathcal{A}_0 \partial_k \Lambda. 
\end{split}
\end{equation}
Therefore the anomalous gauge invariant current is obtained by the compensating shift 
\begin{equation}
 J^\mu_{\mathrm{cov}} = \frac{T_0}{\sqrt{-g}} \frac{\delta W}{\delta \mathcal{A}_\mu}
+ 2 c_A  \epsilon^{\mu \nu \rho \sigma}  \mathcal{A}_\nu 
 \partial_\rho \mathcal{A}_\sigma . 
\end{equation}
Note that the Bardeen-Zumino term in this equation contributes to the current at first order in the derivative expansion.


\subsection{The third order generating functional}
Now we  turn our attention to the connection between    the static  gravitational response at third order in the derivative expansion and the mixed anomaly. 
First, we introduce the Pontryagin density defined by 
\begin{equation}
\mathcal{P} = \frac{1}{2}  {}^\ast R^\mu{}_\nu{}^{\alpha \beta} R^{\nu}{}_{\mu \alpha \beta} , 
\end{equation}
where the dual Riemann tensor is given by  
\begin{equation}
{}^\ast R^\mu{}_\nu{}^{\alpha \beta}  = \frac{1}{2} \epsilon^{\alpha \beta \rho \tau} R^\mu{}_\nu{}_{\rho \tau} . 
\end{equation}
Locally, the Pontryagin density  can be written as a total divergence 
\begin{equation}
\nabla_\mu K^\mu = \mathcal{P} ,
\end{equation}
where $K^\mu$  is the Chern-Simons topological current 
\begin{equation}
K^\alpha = \epsilon^{\alpha \beta \rho \tau} \Gamma^\xi_{\beta \eta}
\Bigl( \partial_\rho \Gamma^\eta_{\tau \xi}  + 
\frac{2}{3}\Gamma^\eta_{\rho \delta} \Gamma^\delta_{\tau \xi}  \Bigr) .
\end{equation}
 
We have seen from \eqref{tenok} that the response to a tensor perturbation of the form \eqref{tensor4} 
is given by 
 \begin{equation}
 \begin{split}\label{tensorial}
 T^{11} & = -T^{22} = 2 c_m \mu h_{\times}'''(z) ,  \\ 
 T^{12} &= -2 c_m \mu h_{+}'''(z)   . 
 \end{split}
 \end{equation}
 This response involving only spatial indices  may actually  be derived from 
 the quadratic portion of the functional\footnote{In the linear approximation the required components read  
 $T^{11} = T_0 \delta W_K^\mathrm{quad}/\delta h_+ = -T^{22}$ and 
  $T^{12} = T_0 \delta W_K^\mathrm{quad}/\delta h_\times$.}  
 \begin{equation} 
 W_K = -\frac{c_m}{T_0} \int d^3 x \sqrt{-g} \mathcal{A}_\mu K^\mu , 
 \end{equation}
 which, for $\mathcal{A}_0=\mu$, $\mathcal{A}_i=0$,  becomes
  \begin{equation}
  W_K^\mathrm{quad}  
  = -\frac{c_m}{T_0}  \int d^3 x \, \mu \bigl( h_{+}'(z)h_{\times}''(z) - h_{+}''(z)h_{\times}'(z) \bigr) .  
  \end{equation}
But   $W_K^\mathrm{quad}$,  when evaluated for the vector perturbation 
  $h_{01}=-a_1(z)$ , $h_{02}=  -a_2(z)$, $h_{03}=0$, 
  \begin{equation}
  W_K^\mathrm{quad} =-\frac{c_m}{T_0}  \int d^3 x \, \mu  K_{\mathrm{quad}}^0 
  = -\frac{c_m}{2T_0}  \int d^3 x \, \mu \bigl(a_2'(z) a_1''(z) - a_1'(z) a_2''(z)\bigr), 
  \end{equation} 
  does not produce the required response. Indeed its variation yields 
 \begin{equation}
 \begin{split}\label{vectorial}
 T^{01} &= -T_0 \frac{\delta W_K^\mathrm{quad}}{\delta a_1} = c_m \mu a_2'''(z) ,  \\ 
 T^{02} &= -T_0 \frac{\delta W_K^\mathrm{quad}}{\delta a_2}  =  -c_m \mu  a_1'''(z) ,  
 \end{split}
 \end{equation}
 which is \emph{minus} our result  \eqref{vecok}. 
 This is consistent with our observations at the end of last section. 
 Therefore, an additional contribution to the action will be  needed in order to properly account for the vector response. 
The following is a natural,  possibly unique choice that   preserves $U(1)$ gauge invariance,  three-dimensional diffeomorphism invariance and 
Kaluza-Klein gauge invariance
\begin{equation}
\label{invar}
W_{\mathrm{inv}} = \frac{c_1}{T_0} \int d^3 x \sqrt{g_3}\,  \epsilon^{i j k} g_{i n} 
\frac{1}{\sqrt{g_3}} \partial_m \bigl(\sqrt{g_3} f^{m n} \bigr) \partial_j \bigl(\mathcal{A}_k - a_k \mathcal{A}_0  \bigr) , 
\end{equation}
where the inverse metric $g^{k l}$ is used to raise the lower indices of $f_{i j} \equiv \partial_i a_j - \partial_j a_i$, 
and $c_1$ is a constant to be determined shortly. 
In terms of differential forms the integrand is proportional to $d(\mathcal{A} - \mathcal{A}_0 a) \wedge \delta d a$, where 
$\delta$ denotes the codifferential. 
The virtue of $W_\mathrm{inv}$ is that  the linearized tensor response 
remains unaffected, while its contribution the vector response  reads
\begin{equation}
\begin{split}
 \label{invc1}
-T_0\frac{\delta W^{(2)}_\mathrm{inv} }{\delta a_1} &=  -2 c_1\mu a_2'''(z) ,  \\ 
-T_0\frac{\delta W^{(2)}_\mathrm{inv} }{\delta a_2} &=  2 c_1   \mu a_1'''(z) . 
 \end{split}
 \end{equation}
 
If we could identify $W_K$ with the anomalous partition function,  then $c_1=c_m$ would be the appropriate choice in order to reproduce our linearized results~\eqref{vecok}. However,   $W_K$ can \textit{not} be identified with the partition function. Instead, the non-invariant functional $W_K$  plays the role of a  local counterterm that interpolates between   
 two alternative definitions of the consistent effective action. 
 These  two choices  preserve  either diffeomorphism or gauge invariance. 
 Following  Bilal's notation~\cite{Bilal:2008qx}, 
if  we denote by $\Gamma^{(1)}$ the diffeormorphism invariant effective action for the mixed gauge-gravitational anomaly, and by  $\Gamma^{(2)}$ the gauge invariant one,  
 the interpolation is given by  
 \begin{equation}
 \label{interpol}
 W_K + \Gamma^{(1)} = \Gamma^{(2)}. 
 \end{equation}
 We must then consider  the contributions to the energy-momentum tensor from $\Gamma^{(1,2)}$ and the relation with
 our results for the static response. 
 
The complete  anomalous contribution to  the partition function  at third derivative order,   
together with its implications on the constitutive relations, 
have been recently stablished in  \cite{Jensen:2012kj}. 
With the vorticity and acceleration of the fluid given by    
\begin{equation}
\begin{split}
\omega^\mu &= \epsilon^{\mu \nu \rho \sigma} u_\nu \nabla_\rho u_\sigma ,  \\
a^\mu&= u^\lambda \nabla_\lambda u^\mu ,
\end{split}
\end{equation}
the authors of  \cite{Jensen:2012kj} construct the well-behaved covariant current  
\begin{equation}
j_m^\mu = -4 W^\mu{}_\nu{}_{\rho \sigma} u^\nu u^\rho \omega^\sigma -
\left( \frac{1}{3} R + 2 R_{\rho \sigma} u^\rho u^\sigma - 2 a^\lambda a_\lambda  - 
\frac{3}{2}\omega^\lambda \omega_\lambda \right) \omega^\mu ,
\end{equation}
where $W^\mu{}_\nu{}_{\rho \sigma}$ is the Weyl tensor  and $R_{\rho \sigma}$ is the Ricci tensor. 
This current, evaluated for 
the equilibrium fluid velocity  $u^\mu = e^{-\sigma} V^\mu$ in the background \eqref{backg}, 
has the same divergence as the topological current
\begin{equation}\label{diver}
\nabla_\mu j_m^\mu = \mathcal{P} .  
\end{equation}
This implies that  the local functionals
\begin{equation}
\begin{split}
\Gamma^{(1)} &= \frac{c_m}{T_0} \int d^3 x \sqrt{-g}\,  \mathcal{A}_\mu j_m^\mu   ,  \\ 
\Gamma^{(2)} &= \frac{c_m}{T_0} \int d^3 x \sqrt{-g}\,  \mathcal{A}_\mu (j_m^\mu - K^\mu)  ,  
\end{split}
\end{equation}
satisfy \eqref{interpol} and have the required properties:  
$\Gamma^{(1)}$ is obviously invariant under three-dimensional diffeomorphisms,  but   not gauge invariant, 
whereas, thanks to~\eqref{diver}, $\Gamma^{(2)}$ is   gauge invariant under $\delta \mathcal{A}_i = \partial_i \Lambda$, $\delta \mathcal{A}_0 = 0$, but not diffeomorphism invariant.  
As a consequence,  these functionals give rise to  two different types of consistent observables. 
 
Let's first consider  the combination $\Gamma^{(2)} + W_\mathrm{inv}$. It is easy to see that the component 
$j_m^0$  is cubic in the metric perturbation  and  does not 
contribute to the linearized $T^{\mu \nu}$. 
 Thus,  for $\mathcal{A}_0$   constant and $\mathcal{A}_i=0$, 
 the quadratic part of $\Gamma^{(2)}$ matches that of $W_K$, 
 and has identical contributions to the linearized $T^{\mu \nu}$. 
 On the other hand, the Bardeen-Zumino term required to compensate for the lack of  invariance under diffeomorphisms   vanishes if the  gauge field strength $\mathcal{F}_{\alpha \beta}$ does~\cite{Jensen:2012kj}, 
and the linear response results 
will agree with the variation 
of the quadratic part of $\Gamma^{(2)} + W_\mathrm{inv}$, 
which coincides with that of $W_K+W_\mathrm{inv}$, for the choice $c_1=c_m$.

For a different  choice of the generating functional,  namely for $\Gamma^{(1)} + W_\mathrm{inv}$, the conclusions are identical, although they require a little more work.  
This functional may be viewed as the third derivative counterpart of 
the first derivative action \eqref{anom1}, since both of them are diffeomorhism invariant. 
Differentiation with respect to $ \mathcal{A}_\mu$ 
yields a gauge and diffeomorfism covariant anomalous current with no need for a compensating term.
But, while the first derivative  term $\delta W/\delta g_{\mu \nu}$ from \eqref{anom1} was gauge invariant, 
the consistent contribution $\delta (\Gamma^{(1)} + W_\mathrm{inv})/\delta g_{\mu \nu}$ is not, 
 and a  Bardeen-Zumino tensor $T^{\mu \nu}_{BZ}$ has to be added 
in order to produce a gauge invariant and generally covariant energy-momentum tensor
\begin{equation}
T^{\mu \nu}= \frac{2 T_0}{\sqrt{-g}} \frac{\delta\bigl(\Gamma^{(1)} + W_\mathrm{inv} \bigr)}{\delta g_{\mu \nu}}  + T^{\mu \nu}_{BZ} . 
\end{equation}

As $j_m^0$  is cubic in the metric perturbation,  $\Gamma^{(1)}$ does not 
contribute to the linearized $T^{\mu \nu}$. Thus  the third-order derivative response,   
which obeys the Ward identity of general covariance, 
must precisely match  the linear portion of the Bardeen-Zumino tensor,  
together with the contribution from $W_\mathrm{inv}$. 
This turns out to be the case if $c_1$ takes the value $c_1=c_m$, which is determined from  \eqref{invc1}. 
Indeed, an explicit computation shows  that  the results in \eqref{vecok} and  \eqref{tenok} 
may be  rewritten as 
\begin{equation}
\delta\langle T^{\mu \nu} \rangle = 2 T_0  \frac{\delta W_\mathrm{inv}^{(2)}}{\delta h_{\mu \nu}}  + 
 2 c_m \partial_j \bigl( 
{}^\ast R^{j \mu \nu  0}_{\mathrm{lin}} \mu_0 + 
{}^\ast R^{j \nu \mu  0}_{\mathrm{lin}}\mu_0 \bigr) .
\end{equation}
The non-linear generalization of the second term,  
\begin{equation}
T^{\mu \nu}_{BZ} =  2 c_m \nabla_\lambda \bigl(  
{}^\ast R^{\lambda \mu \nu \rho} \mathcal{A}_\rho + 
{}^\ast R^{\lambda \nu \mu \rho} \mathcal{A}_\rho \bigr)  , 
\end{equation}
exactly matches  the form of the Bardeen-Zumino energy tensor given in~\cite{Jensen:2012kj} 
when the anomalous piece of the  generating functional is precisely $\Gamma^{(1)}$. 

Thus the results at the third derivative order obtained in this paper 
show that a consistent generating functional  describing the effects of the mixed anomaly 
must include an additional invariant piece $W_\mathrm{inv}$ given by \eqref{invar} with $c_1=c_m$, 
a feature that  ultimately  can be traced to  the opposite signs for   
$c_\mathbb{V}$ and $c_\mathbb{T}$ in~\eqref{c3static}. 


\section{Discussion and outlook}

In this paper we have computed the subleading corrections to  the time-dependent parity violating graviton response function up to third order in the derivative expansion. In particular, we have obtained the complete invariant decomposition of the response function on the appropriate tensor basis~\eqref{tensors6} and~\eqref{tensors4}. We have then exploited  the transversality of the response function to rewrite our results in terms of the two invariant functions $c_\mathbb{V,T}$ which condense all the relevant information.

From the static limit of the response function we have extracted  the  anomalous constitutive relations at first~\eqref{eq:constpv}, and third order~\eqref{vecok}, \eqref{tenok} in the derivative expansion, which we have compared with recent results in the literature. We have also shown that one can reconstruct  the complete parity violating partition function at first order in the derivative expansion~\eqref{anom1} from the linearized corrections to the energy-momentum tensor~\eqref{eq:constpv}. The situation at third order in the derivative expansion is  more involved, but we have shown that our expressions for the constitutive relations are fully compatible with very recent results on the form of the anomalous partition function at that order if one includes the previously unknown parity-odd invariant contribution given by~\eqref{invar}. 

We have also obtained the complete dependence  of the quantities $c_\mathbb{V,T}$  on $q^0/q$, from which one can compute  chiral effects on the time-dependent departures  from equilibrium  $\delta\langle T^{\mu \nu}(t, \bm{q}) \rangle$. 
For the parity-even part of the time-dependent response to metric perturbations, an analysis including  a comparison with the results from the Boltzmann equation has been  given in~\cite{Rebhan:1994zw}. 
Now, having at our disposal a set of results for the parity-odd part  of the response, 
we can  pose the question about the relationship between the field theory approach presented in this paper and
a possible kinetic description involving  the background  metric. 
In this regard, the modification of 
the non-equilibrium kinetic equation that take into  account  chiral magnetic and anomalous Hall effects has been recently
obtained  in~\cite{Son:2012wh,Stephanov:2012ki} in the presence of  an external electromagnetic field. 
In addition to this, 
it would be interesting to take advantage of the results at the linear level in this paper to obtain a Vlasov-type equation 
from which one could derive non-equilibrium chiral effects  caused by metric perturbations or weak curvature backgrounds.
We leave the consideration of this issues for future work\footnote{After this work was completed, we became aware of~\cite{Son:2012zy}, 
which deals with issues similar to the ones mentioned in this paragraph.}.


\acknowledgments
This	work	is supported in part by the Spanish Ministry of Science and Technology under Grant FPA2009-10612 and the Spanish Consolider-Ingenio 2010 Programme CPAN (CSD2007-00042), and by the Basque Government under Grant IT559-10.


\appendix

\section{Expansion of the action and seagull terms}
\label{seag}
In this appendix we obtain eq.~\eqref{sea} for the seagull contribution to the parity-odd part of the response function. 
The action for a fermion in a curved background is given by $S=\int d^4 x\sqrt{-g}\mathcal L$ where
\begin{equation}
\mathcal{L}=\frac{i}{2}\left[\bar\psi\gamma^\mu\nabla_\mu\psi -(\nabla_\mu\bar\psi)\gamma^\mu\psi
\right] ,
\end{equation}
with  $\nabla_\mu\psi=\partial_\mu\psi-\Gamma_\mu\psi$.  
The  spin connection is related to the vierbein $e_a^\nu$ by
\begin{equation}
\Gamma_\mu=\frac{1}{8}[\gamma^a,\gamma^b] e_a^\nu e_{b\nu;\mu}=\frac{1}{8}[\gamma^a,\gamma^b] e_a^\nu (\partial_\mu  e_{b\nu}-\Gamma^\alpha_{\mu\nu}) e_{\beta\alpha} ,
\end{equation}
where greek and latin letters are used for  curved and Minkowski indices respectively, with $\{\gamma^\mu,\gamma^\nu\}=-2g^{\mu\nu}$ and $\{\gamma^a,\gamma^b\}=-2\eta^{ab}$. 
Expanding the action $\mathcal S=\int d^4 x\mathcal L$ in powers of $h_{\mu\nu}=g_{\mu\nu}-\eta_{\mu\nu}$ gives 
$\mathcal{S}=\mathcal{S}_0+\mathcal{S}_1+\mathcal{S}_2+\ldots$ where $\mathcal{S}_0$ is the action in flat space-time, 
$\mathcal{S}_1=\frac{1}{2}\int d^4x \, T^{\mu\nu}h_{\mu\nu}$ with 
\begin{equation}\label{em}
T_{\mu\nu}=\frac{i}{4}\left[\bar\psi\gamma_{\mu}\overleftrightarrow\partial_{\nu}\psi+\bar\psi\gamma_{\nu}\overleftrightarrow\partial_{\mu}\psi\right]-\frac{i}{2}\eta_{\mu\nu}\bar\psi\gamma^{\alpha}\overleftrightarrow\partial_{\alpha}\psi ,
\end{equation}
and
\begin{equation}
\label{s2}
\begin{split}
\mathcal{S}_2&=\frac{i}{16}\int d^4x \bar\psi\left\{ \sigma^{\mu\nu},\gamma^\rho\right\}\psi\, \eta^{\alpha\beta} h_{\alpha\mu}\partial_\rho h_{\beta\nu}\\
&\quad-\frac{1}{8}\int d^4x \left(3\eta^{\alpha\beta}T^{\mu\nu}-2\eta^{\alpha\mu}T^{\beta\nu}\right)h_{\alpha\mu}h_{\beta\nu}\\
&\quad+\frac{1}{8}\int d^4x\left(\eta^{\alpha\beta}\eta^{\mu\nu}-\eta^{\alpha\mu}\eta^{\beta\nu}\right) \mathcal{L}\,h_{\alpha\mu}h_{\beta\nu} ,
\end{split}
\end{equation}
with $\sigma^{\nu\sigma}\equiv\frac{1}{4}[\gamma^\nu,\gamma^\sigma]$. 
The three-vertex fermion-fermion-graviton in eq.~(\ref{v3}) then follows from the Fourier transform of eq.~(\ref{em}). 
Similarly, the seagull contribution in eq.~(\ref{sea}) follows from eq.~(\ref{s2}) through
\begin{equation}\label{seagull}
\begin{split}
\Pi_2^{\mu \nu \, \rho \sigma}(Q)=
-4\left\langle\frac{\delta^2  \mathcal{S}_2}{\delta h_{\mu\nu}\delta h_{\rho\sigma}}\right\rangle 
&=\frac{1}{8}\eta^{\mu\rho}\frac{1}{\beta}\sum_{\omega_n} \int \frac{d^3 k}{(2\pi)^3} \text{tr} 
\bigl[\{\sigma^{\nu\sigma},\slashed{Q}\}\mathcal{P}_-\slashed{K}\bigr]\frac{1}{K^2}  \\ 
&\quad +  \frac{1}{8}\eta^{\nu\rho}\frac{1}{\beta}\sum_{\omega_n} \int \frac{d^3 k}{(2\pi)^3} \text{tr} 
\bigl[\{\sigma^{\mu\sigma},\slashed{Q}\}\mathcal{P}_-\slashed{K}\bigr]\frac{1}{K^2} \\
& \quad+\frac{3}{4}\left(\eta^{\mu\rho}\langle T^{\sigma\nu}\rangle+\eta^{\nu\rho}\langle T^{\sigma\mu}\rangle\right)
+(\rho\leftrightarrow\sigma)\\
&\quad -\left(\eta^{\mu\nu}\langle T^{\rho\sigma}\rangle+\eta^{\rho\sigma}\langle T^{\mu\nu}\rangle\right) ,
\end{split}
\end{equation}
where we have used the fact that the the equations of motion imply $\langle\mathcal{L}\rangle=0$. 
As $\langle T^{\mu\nu}\rangle$  in flat space-time takes the form \eqref{perfect}, which obviously  preserves parity, this establishes the validity of eq.~(\ref{sea}).
 
\section{The integrals $I^{\mu\nu\rho}$}
\label{appi3}
\begin{table}[ht!]
\begin{tabular}{r || c |c| c  }
  & $n=0$ & $n=1$ & $n=2$ \\
\hline \hline
$Q_\alpha Q_\beta Q_\gamma I_{(n)}^{ \alpha\beta \gamma} /I_{(n)}$ & 
$q^0 Q^2 $ & 
$ 3 q^0 Q^4$ &
$0$ \\
$Q_\alpha Q_\beta u_\gamma I_{(n)}^{ \alpha\beta \gamma} /I_{(n)}$ &
$Q^2-\frac{1}{2} q^2$ &
$-3 q^0Q^2$ &
$-6 Q^4 L(q^0, q)$ \\
$Q_\alpha u_\beta u_\gamma I_{(n)}^{ \alpha\beta \gamma} /I_{(n)}$ &
$-q^0$ &
$2q^2-3Q^2 -2Q^2 L(q^0, q)$ &
$12 q^0 Q^2 L(q^0, q)$ \\
$Q_\alpha\eta_{\beta\gamma}  I_{(n)}^{ \alpha\beta \gamma} /I_{(n)} $ &
$q^0$ &
$4Q^2$ &
$0$ \\
$u_\alpha u_\beta u_\gamma I_{(n)}^{ \alpha\beta \gamma} /I_{(n)} $ &
$-1 -\frac{3}{2}L(q^0, q)$ &
$3 q^0 +6 q^0 L(q^0, q)$ &
$2 q^2 -24 (q^0)^2 L(q^0, q)$ \\
$u_\alpha\eta_{\beta\gamma}  I_{(n)}^{ \alpha\beta \gamma} /I_{(n)} $ &
$1$ &
$-4 q^0$ &
$0$ \\
\hline
\end{tabular}
\caption{Contractions  of $I_{(n)}^{ \alpha\beta \gamma}$ with the tensors in eq.~(\ref{tensors6}) for  $n=0,1,2$.}
\label{t1}
\end{table}
Table~\ref{t1} gives the contractions of $I^{\mu\nu\rho}$ with the basis tensors in eq.~(\ref{tensors6}).
The constants $I_{(n)}$  are related to the Fermi distribution as follows
\begin{equation}\label{ints}
\begin{split}
I_{(0)}&= \int_0^\infty \frac{dk}{4\pi^2} k^2 \bigl(n_F(k-\mu) - n_F(k+\mu)\bigr) = \frac{1}{12\pi^2} (\mu^3 + \pi^2 \mu\, T^2),\\
I_{(1)}&= \frac{1}{4}\int_0^\infty \frac{dk}{4\pi^2} k \bigl(n_F(k-\mu) + n_F(k+\mu)\bigr) = \frac{1}{96\pi^2} (3\mu^2 + \pi^2 \, T^2),\\
I_{(2)}&= \frac{1}{48}\int_0^\infty \frac{dk}{4\pi^2}  \bigl(n_F(k-\mu) - n_F(k+\mu)\bigr) = \frac{\mu}{192\pi^2},
\end{split}
\end{equation}
where the prefactors have been chosen for convenience. These relations can be inverted to give the coefficients in the expansion $I^{\mu\nu\rho}=\sum_j c_j T_j^{\mu\nu\rho}$, with the following results:

\begin{itemize}

\item The tensor expansion of $I^{\mu\nu\rho}$ at leading order has been given in eqs.~\eqref{i30} and~\eqref{lq0}.
\item Tensor expansion of $I^{\mu\nu\rho}$ at $\mathcal{O}(Q)$
\begin{equation}
\begin{split}
c_1(q^0, q)/I_{(1)} &= \frac{3}{q^2} + 
          \frac{3 \left(3Q^2-2q^2 \right)}{q^4} L(q^0, q) , \\ 
c_2(q^0, q)/I_{(1)} &=   -\frac{2 q^0}{q^2}  - 
      \frac{6 q^0Q^2}{q^4}  L(q^0, q) , \\ 
c_3(q^0, q)/I_{(1)} &=   -\frac{Q^2}{q^2}  - 
      \frac{3q^0Q^4(5Q^2-2q^2)}{q^6}  L(q^0, q) ,  \\ 
c_4(q^0, q)/I_{(1)} &=   -\frac{Q^2}{q^2}   L(q^0, q) ,  \\ 
c_5(q^0, q)/I_{(1)}    &= 0 , \\
c_6(q^0, q) /I_{(1)}   &= 0  . 
\end{split}
\end{equation}
\item Tensor expansion of $I^{\mu\nu\rho}$ at $\mathcal{O}(Q^2)$
\begin{equation}
\begin{split}A
c_1(q^0, q)/I_{(2)} &= \frac{q^0\left(5Q^2-2q^2\right)}{q^4} + 
          \frac{3 q^0\left(5 Q^4-4q^2Q^2+
          8 q^4  \right)}{q^6} L(q^0, q) , \\ 
c_2(q^0, q)/I_{(2)} &=   \frac{Q^2\left(5Q^2-4q^2\right)}{q^4}  + 
      \frac{3
     Q^2 \left(5 Q^4-6q^2Q^2+
          4 q^4  \right)}{q^6}  L(q^0, q) , \\ 
c_3(q^0, q)/I_{(2)} &=   -\frac{5 q^0Q^4}{q^4}  - 
      \frac{3q^0Q^4(5Q^2-2q^2)}{q^6}  L(q^0, q) ,  \\ 
c_4(q^0, q)/I_{(2)}  &=   -\frac{q^0 Q^2}{q^2}  - 
      \frac{3 q^0Q^4}{q^4}  L(q^0, q) ,  \\ 
c_5(q^0, q)/I_{(2)} &=   -\frac{5 Q^6}{q^4}  - 
      \frac{3 Q^6(5Q^2-2q^2)}{q^6}  L(q^0, q) ,   \\ 
c_6(q^0, q)/I_{(2)} &=   -\frac{Q^4}{q^2}  - 
      \frac{3 Q^6}{q^4}  L(q^0, q) .  
\end{split}
\end{equation}
\end{itemize}
\section{The integrals $I^{\mu\nu}$}
\label{appi2}
Table~\ref{t2} gives the contractions of $I^{\mu\nu}$ with the following basis tensors
\begin{equation}\label{tensors4}
\begin{split}
T^{\mu \nu}_1 &= Q^\mu Q^\nu , \\ 
T^{\mu \nu }_2 &= Q^\mu u^\nu +u^\mu Q^\nu, \\ 
T^{\mu \nu }_3 &= u^\mu u^\nu, \\ 
T^{\mu \nu }_4 &= \eta^{\mu \nu} , \\ 
\end{split}
\end{equation}
with the constants $I_{(n)}$ given in  eq.~(\ref{ints}).

\begin{table}[h!]
\begin{center}
\begin{tabular}{r || c |c }
  & $n=0$ & $n=1$  \\
\hline \hline
$Q_\alpha Q_\beta  I_{(n)}^{ \alpha\beta } /I_{(n+1)}$ &
$-2 Q^2$ &
$0$ \\
$Q_\alpha u_\beta  I_{(n)}^{ \alpha\beta}  /I_{(n+1)}$ &
$2 q^0$ &
$12 Q^2$ \\
$u_\alpha u_\beta  I_{(n)}^{ \alpha\beta}  /I_{(n+1)}$ &
$2+4  L(q^0, q)$ &
$-24 q^0 L(q^0, q)$  \\
$\eta_{\alpha\beta}  I_{(n)}^{ \alpha\beta}  /I_{(n+1)}$ &
$-4$ &
$0$  \\
\hline
\end{tabular}
\caption{Contractions  of $I_{(n)}^{ \alpha\beta }$ with the tensors in eq.~(\ref{tensors4}) for  $n=0,1$.}
\label{t2}
\end{center}
\end{table}
The coefficients of the expansion  $I^{\mu \nu} = \sum_j c_j T^{\mu \nu}_j$ are given by
\begin{itemize}
\item Tensor expansion  of $I^{\mu \nu}$ at leading order 
\begin{equation}
\begin{split}
 c_1(q^0, q)/I_{(1)} &= -\frac{2}{q^2}-\frac{6Q^2-4q^2}{q^4} L(q^0, q) , \\ 
c_2(q^0, q)/I_{(1)} &= \frac{2q^0}{q^2}+\frac{6 q^0Q^2}{q^4} L(q^0, q) , \\ 
 c_3(q^0, q)/I_{(1)} &= \frac{2Q^2}{q^2}+\frac{6Q^4}{q^4} L(q^0, q)  ,  \\ 
c_4(q^0, q)/I_{(1)} &=  \frac{2Q^2}{q^2}  L(q^0, q) .  
      \end{split}
\end{equation}
\item Tensor expansion  of $I^{\mu \nu}$  at $\mathcal{O}(Q)$ 
\begin{equation}
\begin{split}
 c_1(q^0, q)/I_{(2)} &= -24 q^0 L(q^0, q) , \\ 
 c_2(q^0, q)/I_{(2)}  &= -12 Q^2  L(q^0, q) , \\ 
c_3(q^0, q)/I_{(2)}  &= 0 ,  \\ 
 c_4(q^0, q) /I_{(2)}  &=   0 .  
      \end{split}
\end{equation}
\end{itemize}

\bibliographystyle{JHEP}
\bibliography{biboddresponse}

\providecommand{\href}[2]{#2}\begingroup\raggedright\begin{thebibliography}{10}

\bibitem{Son:2009tf}
D.~T. Son and P.~Surowka, {\it {Hydrodynamics with Triangle Anomalies}},  {\em
  Phys.Rev.Lett.} {\bf 103} (2009) 191601,
  [\href{http://xxx.lanl.gov/abs/0906.5044}{{\tt arXiv:0906.5044}}].

\bibitem{Neiman:2010zi}
Y.~Neiman and Y.~Oz, {\it {Relativistic Hydrodynamics with General Anomalous
  Charges}},  {\em JHEP} {\bf 1103} (2011) 023,
  [\href{http://xxx.lanl.gov/abs/1011.5107}{{\tt arXiv:1011.5107}}].

\bibitem{Newman:2005hd}
G.~Newman, {\it {Anomalous hydrodynamics}},  {\em JHEP} {\bf 0601} (2006) 158,
  [\href{http://xxx.lanl.gov/abs/hep-ph/0511236}{{\tt hep-ph/0511236}}].

\bibitem{Kharzeev:2009pj}
D.~E. Kharzeev and H.~J. Warringa, {\it {Chiral Magnetic conductivity}},  {\em
  Phys.Rev.} {\bf D80} (2009) 034028,
  [\href{http://xxx.lanl.gov/abs/0907.5007}{{\tt arXiv:0907.5007}}].

\bibitem{Kharzeev:2010gr}
D.~E. Kharzeev and D.~T. Son, {\it {Testing the chiral magnetic and chiral
  vortical effects in heavy ion collisions}},  {\em Phys.Rev.Lett.} {\bf 106}
  (2011) 062301, [\href{http://xxx.lanl.gov/abs/1010.0038}{{\tt
  arXiv:1010.0038}}].

\bibitem{Landsteiner:2011cp}
K.~Landsteiner, E.~Megias, and F.~Pena-Benitez, {\it {Gravitational Anomaly and
  Transport}},  {\em Phys.Rev.Lett.} {\bf 107} (2011) 021601,
  [\href{http://xxx.lanl.gov/abs/1103.5006}{{\tt arXiv:1103.5006}}].

\bibitem{Jensen:2012jh}
K.~Jensen, M.~Kaminski, P.~Kovtun, R.~Meyer, A.~Ritz, {\em et.~al.}, {\it
  {Towards hydrodynamics without an entropy current}},  {\em Phys.Rev.Lett.}
  {\bf 109} (2012) 101601, [\href{http://xxx.lanl.gov/abs/1203.3556}{{\tt
  arXiv:1203.3556}}].

\bibitem{Jensen:2012jy}
K.~Jensen, {\it {Triangle Anomalies, Thermodynamics, and Hydrodynamics}},  {\em
  Phys.Rev.} {\bf D85} (2012) 125017,
  [\href{http://xxx.lanl.gov/abs/1203.3599}{{\tt arXiv:1203.3599}}].

\bibitem{Banerjee:2012iz}
N.~Banerjee, J.~Bhattacharya, S.~Bhattacharyya, S.~Jain, S.~Minwalla, {\em
  et.~al.}, {\it {Constraints on Fluid Dynamics from Equilibrium Partition
  Functions}},  {\em JHEP} {\bf 1209} (2012) 046,
  [\href{http://xxx.lanl.gov/abs/1203.3544}{{\tt arXiv:1203.3544}}].

\bibitem{Valle:2012em}
M.~Valle, {\it {Hydrodynamics in 1+1 dimensions with gravitational anomalies}},
   {\em JHEP} {\bf 1208} (2012) 113,
  [\href{http://xxx.lanl.gov/abs/1206.1538}{{\tt arXiv:1206.1538}}].

\bibitem{Jensen:2012kj}
K.~Jensen, R.~Loganayagam, and A.~Yarom, {\it {Thermodynamics, gravitational
  anomalies and cones}},  \href{http://xxx.lanl.gov/abs/1207.5824}{{\tt
  arXiv:1207.5824}}.

\bibitem{Bardeen:1984pm}
W.~A. Bardeen and B.~Zumino, {\it {Consistent and Covariant Anomalies in Gauge
  and Gravitational Theories}},  {\em Nucl.Phys.} {\bf B244} (1984) 421.
  Revised version.

\bibitem{Bilal:2008qx}
A.~Bilal, {\it {Lectures on Anomalies}},
  \href{http://xxx.lanl.gov/abs/0802.0634}{{\tt arXiv:0802.0634}}.

\bibitem{Loganayagam:2011mu}
R.~Loganayagam, {\it {Anomaly Induced Transport in Arbitrary Dimensions}},
  \href{http://xxx.lanl.gov/abs/1106.0277}{{\tt arXiv:1106.0277}}.

\bibitem{Loganayagam:2012pz}
R.~Loganayagam and P.~Surowka, {\it {Anomaly/Transport in an Ideal Weyl gas}},
  {\em JHEP} {\bf 1204} (2012) 097,
  [\href{http://xxx.lanl.gov/abs/1201.2812}{{\tt arXiv:1201.2812}}].

\bibitem{Kadanoff}
L.~Kadanoff and P.~C. Martin, {\it Hydrodynamic equations and correlation
  functions},  {\em Annals of Physics} {\bf 24} (1963) 419--469.

\bibitem{Rebhan:1990yr}
A.~Rebhan, {\it {Collective phenomena and instabilities of perturbative quantum
  gravity at nonzero temperature}},  {\em Nucl.Phys.} {\bf B351} (1991)
  706--734.

\bibitem{Tolman30}
R.~C. Tolman and P.~Ehrenfest, {\it Temperature equilibrium in a static
  gravitational field},  {\em Physical Review} {\bf 36} (1930) 1791.

\bibitem{Freedman:1974xm}
D.~Z. Freedman and S.-Y. Pi, {\it {External Gravitational Interactions in
  Quantum Field Theory}},  {\em Annals Phys.} {\bf 91} (1975) 442.

\bibitem{Rebhan:1994zw}
A.~K. Rebhan and D.~J. Schwarz, {\it {Kinetic versus thermal field theory
  approach to cosmological perturbations}},  {\em Phys.Rev.} {\bf D50} (1994)
  2541--2559, [\href{http://xxx.lanl.gov/abs/gr-qc/9403032}{{\tt
  gr-qc/9403032}}].

\bibitem{Son:2012wh}
D.~T. Son and N.~Yamamoto, {\it {Berry Curvature, Triangle Anomalies, and
  Chiral Magnetic Effect in Fermi Liquids}},
  \href{http://xxx.lanl.gov/abs/1203.2697}{{\tt arXiv:1203.2697}}.

\bibitem{Stephanov:2012ki}
M.~Stephanov and Y.~Yin, {\it {Chiral Kinetic Theory}},
  \href{http://xxx.lanl.gov/abs/1207.0747}{{\tt arXiv:1207.0747}}.

\bibitem{Son:2012zy}
D.~T. Son and N.~Yamamoto, {\it {Kinetic theory with Berry curvature from
  quantum field theories}},  \href{http://xxx.lanl.gov/abs/1210.8158}{{\tt
  arXiv:1210.8158}}.

\end{thebibliography}\endgroup

\end{document}